\documentclass[aps,prl,twocolumn,groupedaddress]{revtex4}
\usepackage{graphicx}
\usepackage{gensymb}
\usepackage{amsmath}
\usepackage{amssymb}

\begin{document}

\title{Effects of strain on the electronic structure, superconductivity, and nematicity in FeSe studied by angle-resolved photoemission spectroscopy}
\author{G. N. Phan,$^1$ K. Nakayama,$^1$ K. Sugawara,$^2$ T. Sato,$^1$ T. Urata,$^{1,\ast}$ Y. Tanabe,$^1$ K. Tanigaki,$^{1,2}$ F. Nabeshima,$^3$ Y. Imai,$^{3,\S}$ A. Maeda,$^3$ and T. Takahashi$^{1,2}$}

\affiliation{$^1$Department of Physics, Tohoku University, Sendai 980-8578, Japan\\
$^2$WPI Research Center, Advanced Institute for Materials Research, Tohoku University, Sendai 980-8577, Japan\\
$^3$Department of Basic Science, the University of Tokyo, 3-8-1 Komaba, Meguro, Tokyo 153-8902, Japan}

\date{\today}

\begin{abstract}
One of central issues in iron-based superconductors is the role of structural change to the superconducting transition temperature ($T_c$). It was found in FeSe that the lattice strain leads to a drastic increase in $T_c$, accompanied by suppression of nematic order. By angle-resolved photoemission spectroscopy on tensile- or compressive-strained and strain-free FeSe, we experimentally show that the in-plane strain causes a marked change in the energy overlap ($\varDelta E_{h-e}$) between the hole and electron pockets in the normal state. The change in $\varDelta E_{h-e}$ modifies the Fermi-surface volume, leading to a change in $T_c$. Furthermore, the strength of nematicity is also found to be characterized by $\varDelta E_{h-e}$. These results suggest that the key to understanding the phase diagram is the fermiology and interactions linked to the semimetallic band overlap.
\end{abstract}


\maketitle

\section{I. INTRODUCTION}
Recent advances in fabricating epitaxial thin films have opened a door to utilizing and manipulating the lattice strain by controlling the lattice mismatch with substrate. This strain engineering has a great capability for realizing various physical properties. For example, the epitaxial strain enhances the superconducting transition temperature ($T_c$) in cuprate La$_{2-x}$Sr$_x$CuO$_4$, controls the magnetism and electrical conductivity in manganite La$_{1-x}$Sr$_x$MnO$_3$, induces the room-temperature ferroelectricity in titanate SrTiO$_3$, and realizes a quantum Hall phase in bulk mercury telluride HgTe \cite{Locquet98, Konishi99, Haeni04, Brune11}. Recently a sizable strain effect was reported in iron-based superconductor FeSe \cite {Nabeshima172602, Nie09, Tan13}.

Strain-free bulk FeSe ($T_c$ = 8 K) is a superconductor \cite{Hsu08} which lies in the Bardeen-Cooper-Schrieffer (BCS) and Bose-Einstein condensation (BEC) crossover regime \cite{Kasahara16309}, and exhibits nematic order \cite{Nakayama237001, Shimojima121111, Watson155106, Zhang214503, Huynh144516, Baek15, Bohmer027001} where the C$_4$ rotational symmetry in the electronic structure is broken. It was reported that compressive strain enhances the $T_c$ by 50$\%$ \cite{Nabeshima172602}, while tensile strain kills the superconductivity \cite{Nie09} and simultaneously enhances the nematicity \cite{Tan13}. A question naturally arises as to what triggers such characteristic strain effects. Elucidating this issue would provide a clue to understanding the origins of superconductivity and nematicity as well as their interplay, which are at the heart of intensive investigations of iron-based superconductors. Since the low-energy electronic structure, which governs the key electronic properties, is sensitive to lattice strain \cite{Dhaka14, Gofryk14}, the experimental determination of the electronic structure in strained FeSe is of crucial importance. However, such an experiment has been limited to the nematic phase of the tensile-strained nonsuperconducting regime \cite{Tan13}.

In this article, we traced the evolution of electronic structure with epitaxial strain by angle-resolved photoemission spectroscopy (ARPES) on FeSe. We fabricated a compressive-strained FeSe film on CaF$_2$ substrate (FeSe/CaF$_2$; $T_c$ = 12 K) and a tensile-strained film on SrTiO$_3$ (FeSe/SrTiO$_3$; nonsuperconducting), and compared their electronic structure with that of strain-free bulk FeSe ($T_c$ = 8 K).

 \begin{figure*}
 \includegraphics[width=6.4in]{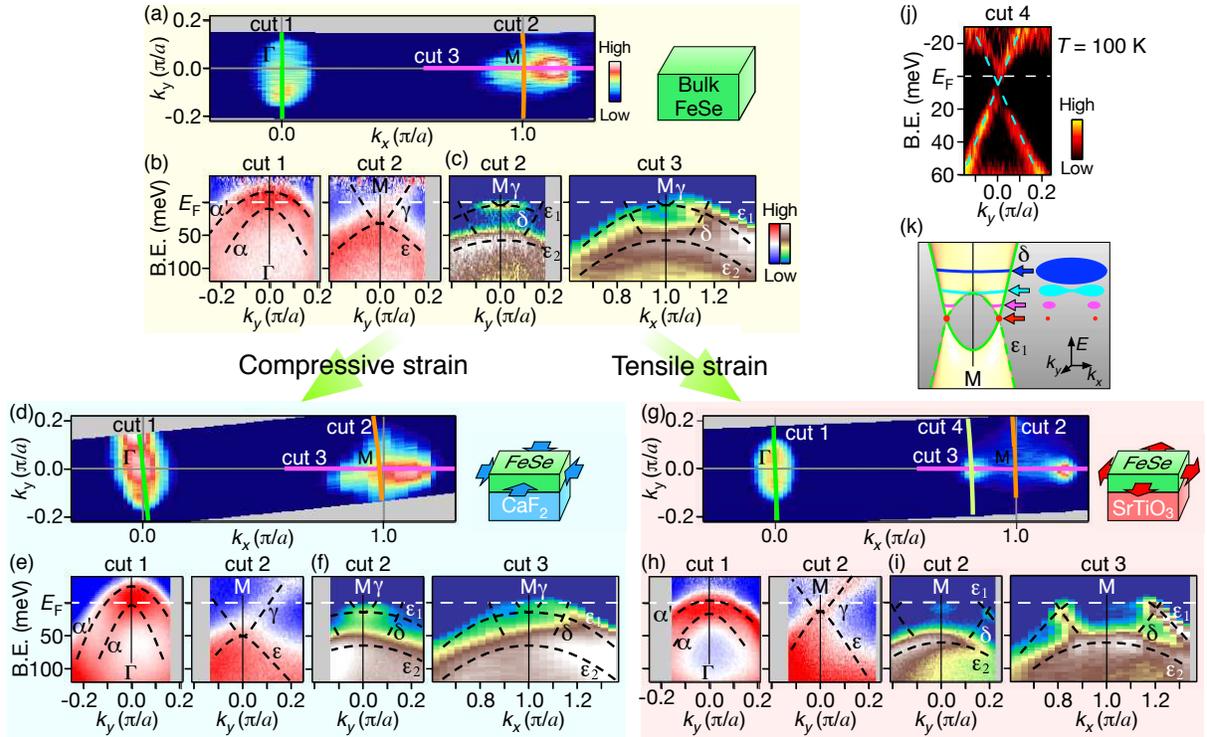}
\caption{(color online) Comparison of electronic structure among bulk FeSe, FeSe/CaF$_2$, and FeSe/SrTiO$_3$. (a) ARPES intensity map at $E_F$ as a function of two-dimensional wave vector measured in the nematic phase ({\it T }= 30 K) for strain-free bulk FeSe ($T_c$ $\sim$ 8 K). The intensity was obtained by integrating the spectral intensity within $\pm $10 meV with respect to $E_F$. (b) Near-$E_F$ ARPES intensity plot for bulk FeSe as a function of binding energy and wave vector measured in the normal state ({\it T} = 120 K) along cuts 1 and 2 shown in (a). The data were divided by the Fermi-Dirac function convoluted with the instrumental resolution. Dashed curves are a guide for the eyes to trace the band dispersions. (c) Near-$E_F$ ARPES intensity measured at {\it T} = 30 K along cuts 2 and 3 in (a). (d)-(f) and (g)-(i) Same as (a)-(c), but for compressive-strained FeSe/CaF$_2$ ($T_c$ $\sim$ 12 K) and tensile-strained FeSe/SrTiO$_3$ ($T_c$ $\sim$ 0 K), respectively. The ARPES intensities in (e) and (h) were measured at {\it T} = 120 K and 180 K, respectively. (j) Second-derivative intensity plot measured along cut 4 in (g) at {\it T} = 100 K, obtained after dividing by the Fermi-Dirac function convoluted with the instrumental resolution. Blue dashed lines are a guide for the eyes to trace the Dirac-cone-like dispersion. (k) Schematic illustration of Dirac-cone dispersion around the {\it M} point. Blue, light blue, pink, and red arrows highlight the Lifshitz transition of FS. The former two highlight the case of FeSe/SrTiO$_3$, while the latter two the case of bulk FeSe and FeSe/CaF$_2$.}
\end{figure*}

\section{II. METHODS}
High-quality FeSe thin films were grown on CaF$_2$ and SrTiO$_3$ substrates by pulsed-laser-deposition and molecular-beam-epitaxy methods, respectively. The film thicknesses of FeSe/CaF$_2$ ($T_c$ = 12 K) and FeSe/SrTiO$_3$ (non-superconducting) samples are approximately 300 monolayers (MLs) and 20 MLs, respectively. These values are much larger than the critical thickness of 8 MLs, above which the finite size effect is negligible \cite{Song020503}. It is also noted that the tensile strain is not relaxed at the top surface of 20-ML FeSe/SrTiO$_3$ as demonstrated by the thickness dependence study \cite{Tan13}. For FeSe/CaF$_2$, it has been reported that the compressive strain is applied to relatively thick films but not to ultrathin films because of the Volmer-Weber-type growth; the reported in-plane lattice constant at $\sim$100 MLs or less is close to that of bulk FeSe, while it becomes smaller by $\sim$1\% at $\sim$150 MLs and this compressive strain does not relax at least until 400 MLs \cite{Nabeshima172602}. By taking into account the thickness independence of the compressive strain over a wide range between 150 and 400 MLs, we have chosen the thickness of 300 MLs to apply compressive strain in the present study. The absence of lattice relaxation at the top surface of our 20-ML FeSe/SrTiO$_3$ and 300-ML FeSe/CaF$_2$ is confirmed by the in-plane lattice constant estimated from the Brillouin-zone size determined by the present ARPES experiments (see Supplemental Material for details of the estimation of the lattice constant by ARPES \cite{Suppl}). These ensure the observed differences between thin films and bulk crystals to be essentially of strain-effect origin. It is also noted that all the samples have been kept non-doped to avoid possible complications from the carrier doping; in this regard, they are distinct from heavily electron-doped FeSe/SrTiO$_3$ \cite{Wang037402, Miyata15, Wen16}. High-quality bulk single crystals ($T_c$ = 8 K) were grown by the KCl and AlCl$_3$ flux method. Details of the sample preparation are described elsewhere \cite{Nabeshima172602, Huynh144516, Miyata15}.
 
ARPES measurements were performed with a Scienta-Omicron SES2002 electron analyzer with a He discharge lamp ($h\nu$ = 21.218 eV) at Tohoku University. The energy and angular resolutions were set at 12-30 meV and 0.2$^{\circ}$, respectively. The Fermi level ($E_F$) of the samples was referenced to that of a gold film evaporated onto the sample holder.
 
 \section{III. RESULTS and DISCUSSION}
Figure 1 displays a comparison of Fermi surface (FS) and representative band dispersions for three momentum cuts among strain-free bulk FeSe [Figs. 1(a)-1(c)], compressive-strained FeSe/CaF$_2$ [Figs. 1(d)-1(f)], and tensile-strained FeSe/SrTiO$_3$ [Figs. 1(g)-1(i)]. In the normal state of bulk FeSe, one can recognize a holelike FS originating from a highly dispersive band ($\alpha'$) around the $\Gamma$ point in the Brillouin zone [cut 1 of Fig. 1(b)]. Another holelike band ($\alpha$) with the top of dispersion below the Fermi level ($E_F$) is also recognized. These bands persist in the nematic phase, keeping their energy positions nearly stationary to temperature \cite{Nakayama237001, Shimojima121111, Watson155106, Zhang214503}. On the other hand, the band structure around the {\it M} point is significantly reconstructed by nematicity. In the normal state [cut 2 of Fig. 1(b)], there exist a shallow electronlike band ($\gamma$) and a holelike band ($\varepsilon$), which degenerate at {\it M} to produce a van Hove singularity in the band structure \cite{Nakayama237001}. In the nematic phase [Fig. 1(c)], the $\varepsilon$ band splits into two branches ($\varepsilon_1$ and $\varepsilon_2$) due to the lifting of $d_{xz}/d_{yz}$-orbital degeneracy, and simultaneously the $\gamma$ band shifts upward \cite{Nakayama237001, Shimojima121111, Watson155106, Zhang214503}. As shown in Figs. 1(d)-1(i), the overall band structure of compressive-strained FeSe/CaF$_2$ and tensile-strained FeSe/SrTiO$_3$ is qualitatively similar to that of the bulk counterpart; one can recognize the semimetallic nature of the band structure regardless of temperature as well as a finite splitting between the $\varepsilon_1$ and $\varepsilon_2$ bands at low temperature, indicative of the commonality of the nematic phase for strain-free and strained FeSe. A closer look at Fig. 1 further reveals some quantitative differences in the Fermi wave vector of the $\alpha'$ band [cut 1 of Figs. 1(b), 1(e), and 1(h)], the position of the van Hove singularity [cut 2 of Figs. 1(b), 1(e), and 1(h)], and the location of the $\gamma$, $\varepsilon_1$, and $\varepsilon_2$ bands in the nematic phase [Figs. 1(c), 1(f), and 1(i)].
 
Another important consequence of the epitaxial strain is the change in the FS topology in the nematic phase. In FeSe/SrTiO$_3$, the electron pocket at {\it M} is absent [Fig. 1(g)] unlike FeSe/CaF$_2$ and bulk FeSe [Figs. 1(a) and 1(d)], and small electron pockets emerge away from {\it M}. This Lifshitz transition is associated with the Dirac-cone-like band dispersion \cite{Kuroki09, Pandey12}, as highlighted by the ``X''-shaped dispersion in FeSe/SrTiO$_3$ [Fig. 1(j), see also Supplemental Material Fig. S2]. As shown in Fig. 1(k), there exist two Dirac cones originating from the hybridization of the $\varepsilon_1$ and $\delta$ bands (note that the intensity of the $\delta$ band is too faint to resolve at high temperatures). When $E_F$ is below the $\varepsilon_1$-band top (case of FeSe/SrTiO$_3$), each Dirac cone creates a small FS pocket slightly away from {\it M}. On the other hand, when $E_F$ is above the $\varepsilon_1$-band top (cases of bulk FeSe and FeSe/CaF$_2$), two Dirac cones are connected with each other to form a single FS.

\begin{figure}
 \includegraphics[width=3.4in]{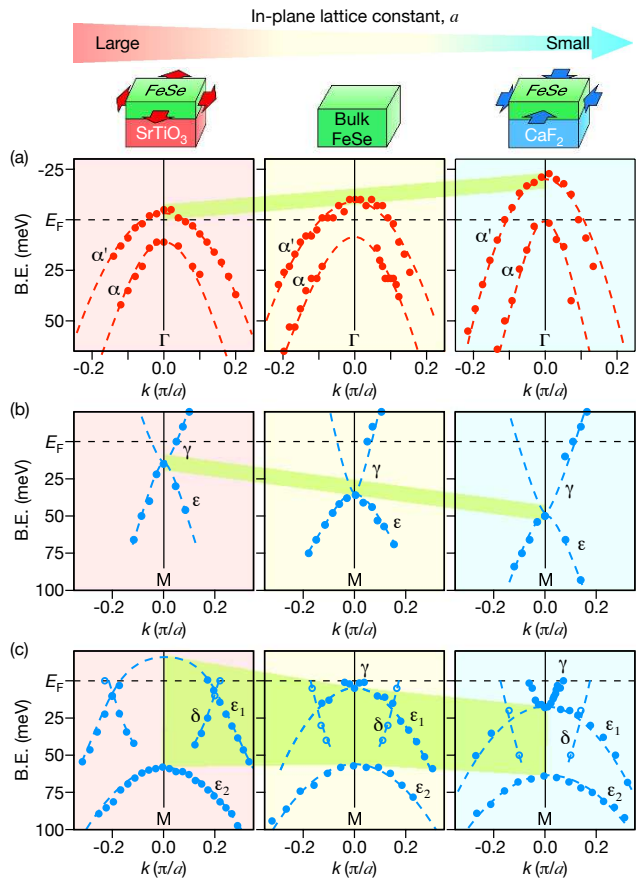}
\caption{(color online) Strain-induced evolution of band dispersions. (a) Comparison of near-$E_F$ band dispersions around the $\Gamma$ point in the normal state ({\it T} = 120-190 K) for FeSe/SrTiO$_3$, bulk FeSe, and FeSe/CaF$_2$. (b) and (c), Same as (a), but measured around the {\it M} point at high temperature ({\it T} = 120-190 K) and low temperature ({\it T} = 30 K), respectively. Filled circles were extracted from the peak maxima in ARPES spectra shown in Fig. 1. Open circles for the $\delta$ band were extracted from the second-derivative plots in supplementary Fig. S2. Dashed curves are a guide for the eyes to trace the band dispersions. Green shade in (a)-(c) highlights the top of the $\alpha'$ band, the bottom of the $\gamma$ band, and the energy scale of $\varDelta E_{nem}$, respectively.}
\end{figure}

To see the strain-induced evolution quantitatively, we show in Fig. 2(a) a side-by-side comparison of the band dispersion around $\Gamma$. Obviously, the $\alpha'$ band shifts {\it upward} with decreasing the in-plane lattice constant {\it a}. In contrast, at {\it M} [Fig. 2(b)], the $\gamma$ and $\varepsilon$ bands shift {\it downward} with decreasing {\it a}. Such an opposite band shift indicates that the energy overlap between the $\alpha'$ and $\gamma$ bands in the normal state, defined here as semimetallic band overlap $\varDelta E_{h-e}$, strongly depends on the strength of strain. In fact, $\varDelta E_{h-e}$ is $\sim$70 meV for FeSe/CaF$_2$ while it is $\sim$20 meV for FeSe/SrTiO$_3$. The observed change in $\varDelta E_{h-e}$ is not due to the $k_z$ dispersion but due to the relative energy shift between the $\alpha'$ and $\gamma$ bands \cite{note}. This demonstrates that the increase of tensile strain significantly reduces both the hole- and electron-carrier concentrations while keeping the semimetallicity. As recognized from Fig. 2(c), an energy separation of the $\varepsilon_1$ and $\varepsilon_2$ bands in the nematic state ($\varDelta E_{nem}$), which reflects the strength of nematicity \cite{Nakayama237001, Shimojima121111, Watson155106, Zhang214503}, is enhanced from $\sim$45 meV in FeSe/CaF$_2$ to $>$70 meV in FeSe/SrTiO$_3$ upon increasing {\it a}. This establishes the anti-correlation between nematicity and superconductivity.

\begin{figure}
 \includegraphics[width=3.4in]{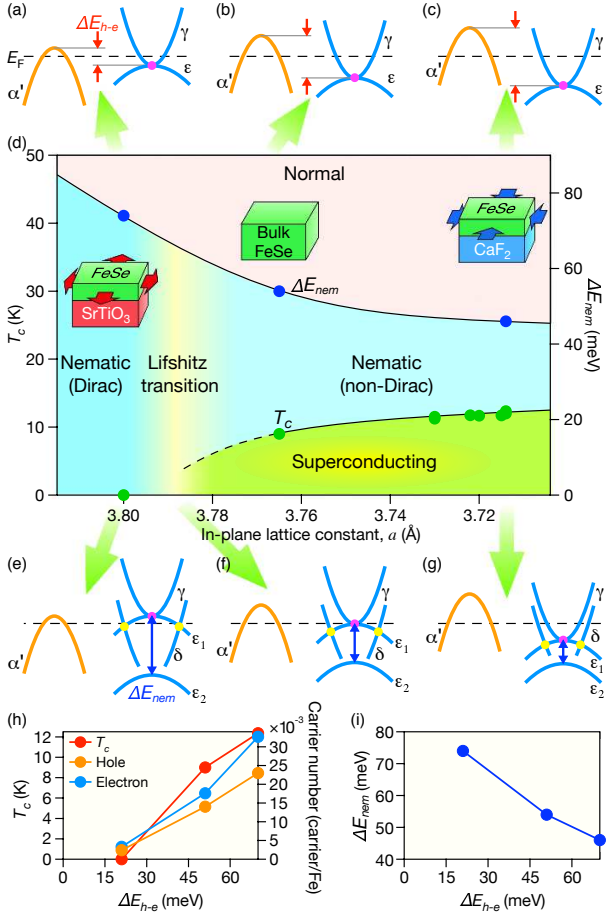}
 \caption{(color online) Relationship between semimetallic band gap and some key physical parameters. (a)-(c) Schematics of normal-state band structure for tensile-strained, strain-free, and compressive-strained FeSe, respectively. Pink dots indicate the position of van Hove singularity at the {\it M} point, while red allows highlight the magnitude of $\varDelta E_{h-e}$, which is the energy difference between the $\alpha'$-band top and the $\gamma$-band bottom. (d) Electronic phase diagram of FeSe as a function of in-plane lattice constant {\it a} \cite{Nabeshima172602, Tan13, Huynh144516}. (e)-(g) Schematics of the band structure in the nematic state for tensile-strained phase, at a critical point of the Lifshitz transition, and for compressive-strained phase, respectively. Yellow dots indicate the position of the Dirac point. (h) Strain dependence of $T_c$ and hole/electron carrier concentration at {\it T} = 30 K plotted as a function of $\varDelta E_{h-e}$. (i) Plot of $\varDelta E_{nem}$ (energy difference between the $\varepsilon_1$ and $\varepsilon_2$ bands at the {\it M} point) as a function of $\varDelta E_{h-e}$.}
\end{figure}

Having established the strain-induced band evolution, the next question is how it is related to superconductivity and nematicity. We propose that these electronic phases are sensitive to the semimetallic band overlap $\varDelta E_{h-e}$. As illustrated in the schematic normal-state band dispersion in Figs. 3(a)-3(c), the $\varDelta E_{h-e}$ gradually increases with decreasing {\it a}. When we estimate the electron- and hole-carrier concentrations from the FS volume in the nematic phase, they have a positive correlation to $\varDelta E_{h-e}$ [Fig. 3(h)]. Intriguingly, the $T_c$ also has a positive correlation, indicating that a larger $\varDelta E_{h-e}$ which results in a larger carrier concentration may be essential for achieving higher $T_c$ in compensated FeSe.
 
Next we discuss how $\varDelta E_{h-e}$ is related to the strength of nematicity. The small $\varDelta E_{h-e}$ under tensile strain means that the van Hove singularity with high density of states \cite{Nakayama237001} is located close to $E_F$ in the normal state [red dot in Fig. 3(a)]. In this case, the band splitting at {\it M} is highly promoted in the nematic phase, because the larger splitting between the $\varepsilon_1$ and $\varepsilon_2$ bands (i.e., larger $\varDelta E_{nem}$) and the resultant $E_F$-crossing of the $\varepsilon_1$ band [Fig. 3(e)] would cause a larger energy gain to stabilize the nematic state. This is inferred from the negative correlation of $\varDelta E_{nem}$ to $\varDelta E_{h-e}$ [Fig. 3(i)], which is also consistent with a recent theoretical study reporting the Pomeranchuk-type nematic instability at $\varDelta E_{h-e}$ $\sim$ 0 \cite{Jiang15}.
 
The Lifshitz transition is also related to $\varDelta E_{h-e}$ through the energy balance with $\varDelta E_{nem}$. Due to the anti-correlation between $\varDelta E_{h-e}$ and $\varDelta E_{nem}$ [Fig. 3(i)], there exists a Lifshitz point at which the $\varepsilon_1$-band top just touches $E_F$ in the nematic phase [Fig. 3(f)]. When the $\varepsilon_1$-band top is further pushed upward into the unoccupied region [tensile-strained case; Fig. 3(e)], the Dirac cones are pushed upward as well. Consequently, the Dirac points are situated in the vicinity of $E_F$. In this regard, the nematic phase could be categorized into two regimes: one with Dirac carriers in the tensile-strained region, and the other without Dirac carriers in the strain-free and compressive-strained regions [Fig. 3(d)]. Our observation thus suggests that the electronic properties of the nematic phase can be sensitive to $\varDelta E_{h-e}$, which would make strained FeSe a promising platform to explore unconventional physical properties arising from Dirac carriers.
 
While the change in carrier concentration may be essential to the $T_c$ variation [Fig. 3(h)], the Lifshitz transition may also play a role in the suppression of superconductivity. It has been suggested that spin fluctuations are involved in the superconducting pairing in bulk FeSe due to the existence of gap nodes \cite{Kasahara16309, Song1410}. While spin fluctuations promote the pair scattering between the hole and electron pockets via {\it Q} = ($\pi$, 0) \cite{Mazin057003, Kuroki087004}, such a scattering can be significantly suppressed in tensile-strained FeSe due to the poorly-defined {\it Q} = ($\pi$, 0) scattering originating from the shrinkage of the hole pocket at $\Gamma$ and the emergence of Dirac-cone-induced small pockets away from {\it M} [Fig. 1(g)]. It is noted that a similar Lifshitz transition around the Brillouin-zone corner has been reported to coincide with the onset of superconductivity in lightly-doped BaFe$_{2-x}$Co$_x$As$_2$ \cite{Liu10}. This supports a close correlation between the observed Lifshitz transition and the suppression of superconductivity under tensile strain.
 
Our observation also has an important implication to the BCS-BEC crossover \cite{Kasahara16309}. A key parameter of this crossover is an effective Fermi energy ($\varepsilon_F$), an energy separation between $E_F$ and the top (bottom) of hole (electron) pocket. The relatively large $k_{B}T_{c}/\varepsilon_F$ ratio of $\sim$0.2 in bulk FeSe indicates that strain-free FeSe lies in this crossover regime \cite{Kasahara16309}. We found that an increase of $\varepsilon_F$ under compressive strain reduces $k_{B}T_{c}/\varepsilon_F$ to 0.05-0.07 in FeSe/CaF$_2$. This suggests that $T_c$ increases on approaching the conventional BCS side, signifying that the proximity of the BCS-BEC crossover, which could be a possible route to high-$T_c$ superconductivity \cite{Randeria14}, is unlikely to be essential for achieving higher $T_c$ in FeSe. Our result demonstrates that $\varepsilon_F$ can be tuned via the epitaxial strain through the control of $\varDelta E_{h-e}$ and $\varDelta E_{nem}$. Such tunability provides a useful route to investigating the BCS-BEC crossover physics.

Finally, we discuss the origin of the strain-induced band evolution and its implications. The conventional band calculations do not correctly reproduce the magnitude of experimental $\varDelta E_{h-e}$ even in strain-free FeSe; namely $\varDelta E_{h-e}$ is significantly overestimated, as discussed in a previous study \cite{Terashima14}. One possible clue to reconcile with the experiment would be the electronic correlation effect, e.g., the repulsive interband interactions which have been proposed to reduce $\varDelta E_{h-e}$ \cite{Ortenzi09, Hirschfeld11}. In the present ARPES, $\varDelta E_{h-e}$ decreases as the effective mass $m^*/m_e$ of the $\alpha'$ band increases ($m_e$ is the free-electron mass and $m^*/m_e$ varies from 3.3 in FeSe/CaF$_2$ to 3.8 in FeSe/SrTiO$_3$). This suggests a certain role of high-energy interactions in reducing $\varDelta E_{h-e}$. The strain-induced change in the electronic correlation may also play an important role in controlling the electronic phase diagram, since it has been suggested that superconductivity and nematicity are sensitive to the high-energy interactions as well as the low-energy electronic structure in iron-based superconductors due to the moderately correlated nature. In fact, the change in $\varDelta E_{h-e}$ alone cannot explain the strong material dependence of $T_c$, and hence the change in the strength of pairing interactions should be taken into account.

\section{IV. SUMMARY}
We reported high-resolution ARPES results on epitaxial films and bulk crystals of FeSe to elucidate the evolution of the band structure from the tensile-strained non-superconducting regime to the compressive-strained high-$T_c$ regime. We found that the compressive strain gives rise to large semimetallic band overlap $\varDelta E_{h-e}$ and the resultant expansion of hole and electron Fermi surfaces, leading to higher $T_c$. Under tensile strain, small $\varDelta E_{h-e}$ results in the shrinkage of Fermi surfaces and the enhancement of nematic instability, leading to the suppression of superconductivity accompanied by the stabilization of nematicity. We also observed the reduction of mass renormalization factor under compressive strain due to the weakening of electronic correlation, which may be also responsible for large $\varDelta E_{h-e}$ and higher $T_c$. Our finding related to the semimetallic band overlap provides important insights into the origins of superconductivity and nematicity as well as their interplay in iron-based superconductors.

\section{ACKNOWLEDGEMENTS}
We thank Y. Miyata, S. Kanayama, N. Inami, H. Kumigashira, and K. Ono for their assistance in ARPES measurements. This work was supported by grants from the Japan Society for the Promotion of Science (JSPS), KAKENHI (grants No. JP15H02105, No. JP25287079, No. JP15H05853, and No. JP25107003), the Program for Key Interdisciplinary Research, the Murata Science Foundation, and High Energy Accelerator Research Organization, Photon Factory (KEK-PF) (Proposals No. 2012S2-001 and No. 2015S2-003).

%\clearpage

\bibliographystyle{prsty}

\begin{thebibliography}{30}
\bibitem{note} It is reported that the $\alpha'$ band shows a finite $k_z$ dispersion in bulk FeSe \cite{Watson155106, Zhang214503}. If the longer (shorter) $c$-axis length in FeSe/CaF$_2$ (FeSe/SrTiO$_3$) dramatically reduces (enhances) the $k_z$ dispersion, the $\alpha'$ band would be shifted downward (upward) since the photon energy of 21.218 eV probes $k_z \sim \pi$ where the $\alpha'$-band top shows the local maximum. Since our ARPES data obviously show an opposite trend, one can conclude that the observed band shift is not dominated by the change in $k_z$ dispersion.
\end{thebibliography}

\end{document}